\documentstyle[12pt,aps,epsf]{revtex}
\def\s{\sigma}

\def\up{\uparrow}
\def\dd{\downarrow}
\def\las{\langle}
\def\ras{\rangle}

\def\nn{\nonumber}

\begin{document}

\title{The effects of a magnetic barrier and a nonmagnetic spacer in
tunnel structures}
\author{Ali A. Shokri$^1$ \footnote{E-mail: aashokri@mehr.sharif.edu} and Alireza
Saffarzadeh$^{2}$
\footnote{Corresponding author. E-mail: a-saffar@tehran.pnu.ac.ir}\\
$^1$ Department of Physics, Sharif University of Technology, 11365-9161, Tehran, Iran \\
$^2$ Department of Physics, Tehran Payame Noor University,
Fallahpour St., Nejatollahi St., Tehran , Iran}
\date{\today}
\maketitle

\begin{abstract}
The spin-polarized transport is investigated in a new type of magnetic
tunnel junction which consists of two ferromagnetic electrodes separated
by a magnetic barrier and a nonmagnetic metallic spacer. Based on
the transfer matrix method and the nearly-free-electron-approximation the
dependence of the tunnel magnetoresistance (TMR) and electron-spin
polarization on the nonmagnetic layer thickness and the applied bias voltage
are studied theoretically. The TMR and spin polarization show an oscillatory
behavior as a function of the spacer thickness and the bias voltage.
The oscillations originate from the quantum well states in the
spacer, while the existence of the magnetic barrier gives rise to a strong
spin polarization and high values of the TMR. Our results may be useful for
the development of spin electronic devices based on coherent transport.
\end{abstract}
\newpage
\section{Introduction}
During recent years, the large magnetoresistance of magnetic
tunnel junctions (MTJs) has attracted much attention due to the
possibility of its application in digital storage and magnetic
sensor technologies \cite{Mood1,Daugh}. In most of the
experiments, the MTJs consist of two ferromagnetic metal (FM)
electrodes separated by a thin insulator (FM/I/FM). In these
structures, the electrical resistance is reduced when the
magnetization direction of the FM layers changes from antiparallel
to parallel alignment by an external magnetic field. Because the
tunnel magnetoresistance (TMR) strongly depends on the spin
polarization of the magnetic electrodes \cite{Jul,Mae}, attempts
have been made to fabricate junctions with more highly
spin-polarized ferromagnets \cite{Lu,Sen}.

Recent experiments on the TMR have shown that the insertion of a thin
nonmagnetic metal (NM) layer between insulating barrier and the FM
electrode (FM/I/NM/FM) reduces the TMR, when thickness of the NM
layer increases \cite{Mood2,Sun,Lec1,Lec2}. This decrease of the TMR can be
attributed to the decoherence tunneling of electrons from one electrode to
the other through the NM layer \cite{Zhang}. On the other hand, recent
observations by Yuasa {\it et al} \cite{Yuasa} show clear oscillations of the
TMR in high-quality NiFe/Al$_2$O$_3$/Cu/Co junctions in which the Cu/Co is a
single crystal. This oscillatory behavior has been interpreted in terms
of the formation of spin-polarized resonant tunneling.

The effect of nonmagnetic metallic interlayers has been
theoretically investigated in FM/NM/I/NM/FM structures. Based on
the Kubo formalism, Vedyayev $et$ $al.$ \cite{Ved} studied
magnetoconductivity as a function of NM layer thickness at low
bias voltages. By using the transfer matrix approach, Wilczynski
and Barnas \cite{WilBar} calculated the thickness and voltage
dependence of TMR. These calculations showed that the presence of
thin NM spacers can lead to the formation of quantum well states
that lead to oscillations of magnetoconductivity and TMR. When the
thickness of only one of the NM layer varies, and also in the case
of an asymmetric structure, such as FM/NM/I/FM \cite{Zhang2}, the
sign of magnetoconductivity and TMR is predicted to oscillate as a
function of the NM layer thickness.

TMR has also been investigated both experimentally
\cite{Lec3,Lec4} and theoretically \cite{Worl,Wil,Saffar1,Saffar2}
in spin filter tunnel junctions in which a ferromagnetic
semiconductor (FMS) is applied as a magnetic barrier. The large
spin polarization achievable using magnetic barriers \cite{Mood3}
makes spin filtering a nearly ideal method for spin injection into
semiconductors, enabling novel spintronic devices \cite{Fied}.
When an FMS layer which acts as a spin filter is used in tunnel
structures, due to the spin splitting of its conduction band,
tunneling electrons see a spin-dependent barrier height. Thus one
spin channel will have a larger tunneling probability than the
other, and a highly spin-polarized current may result.

The purpose of the present work is to investigate theoretically
the effect of a magnetic barrier on spin transport in the presence
of a nonmagnetic spacer layer. Using the transfer matrix method
and the nearly-free-electron approximation we will study the
variation of TMR and spin polarization of tunneling electrons in
terms of the spacer thickness and bias voltage at $T$=0 K. We
assume that the electron wave vector parallel to the interfaces
and the electron spin are conserved in the tunneling process
through the whole system.

The paper is organized as follows. In section 2, the model and the
analytic formulae of the spin polarization and TMR for
FM/FMS/NM/FM structure are presented. In section 3, numerical
results obtained for a typical tunnel junction are discussed. The
results of this work are summarized in section 4.

\section{Description of model and formalism}
We consider two semi-infinite FM electrodes separated by a
ferromagnetic barrier which acts as a spin filter, and a
nonmagnetic metallic layer as a quantum well, in the presence of
DC applied bias $V_a$ as shown in Fig. 1. For simplicity, we
assume that the two FM electrodes are made of the same metal and
all of the interfaces are flat. In a nearly-free-electron
approximation of spin-polarized conduction electrons, the
longitudinal part of the effective one-electron Hamiltonian can be
written as
\begin{equation}\label{H}
H_x=-\frac{\hbar^2}{2m_j^*}\frac{d^2}{dx^2}+U_j(x)-{\bf h}_j\cdot{\bf\s}
+V^\s\ ,
\end{equation}
where $m^*_j$ ($j$=1-4) is the electron effective mass in the $j$th layer, and
\begin{equation}\label{U}
U_j(x)=\left\{\begin{array}{cc}
0, & x<0 \ ,\\
E_{\rm FL}+\phi-eV_ax/t_{\rm bar}, & 0<x<t_{\rm bar}\  ,\\
-eV_a, & t_{\rm bar}<x<t_{\rm bar}+t_{\rm NM}\  ,\\
-eV_a, & x>t_{\rm bar}+t_{\rm NM}\  ,\\
\end{array}\right.
\end{equation}
where $E_{\rm FL}$ is the Fermi energy in the left electrode.
$-{\bf h}_j\cdot{\bf\s}$ is the internal exchange energy where
${\bf h}_j$ is the molecular field in the $j$th FM electrode and ${\bf\s}$ is
the conventional Pauli spin operator.
The last term in Eq. (\ref{H}) is a spin-dependent potential and
denotes the $s-f$ exchange coupling between the spin of tunneling electrons
and the localized $f$ spins in the magnetic barrier. This term, within the
mean-field approximation, is proportional to the thermal average of the $f$
spins, $\las S_z\ras$, and can be written as $V^\s=-\s I\las S_z\ras$.
Here, $\s=\pm 1$ which corresponds to $\s=\up,\dd$, and  $I$ is the $s-f$
exchange constant in the magnetic barrier.

The Schr\"odinger equation for a biased barrier layer can be
simplified by a coordinate transformation whose solution is the
linear combination of the Airy function Ai[$\rho(x)$] and its
complement Bi[$\rho(x)$] \cite{Abram}. Considering all four
regions of the FM/FMS/NM/FM junction shown in Fig. 1, the
eigenfunctions of the Hamiltonian (\ref{H}) with eigenvalue $E_x$
have the following forms:
\begin{equation}\label{psi}
\psi_{j,\s}(x)=\left\{\begin{array}{cc}
A_{1\s}e^{ik_{1\s}x}+B_{1\s}e^{-ik_{1\s}x}, & x<0 \ ,\\
A_{2\s}{\rm Ai}[\rho_{\s}(x)]+B_{2\s}{\rm Bi}[\rho_{\s}(x)], &
0<x<t_{\rm bar}\ ,\\
A_{3\sigma}e^{ik_{3}x}+B_{3\s}e^{-ik_{3}x}, & t_{\rm bar}<x<t_{\rm bar}+
t_{\rm NM}\  ,\\
A_{4\s}e^{ik_{4\s}x}+B_{4\s}e^{-ik_{4\s}x},& x>t_{\rm bar}+t_{\rm NM}\ ,\\
\end{array}\right.
\end{equation}
where,
\begin{equation}
k_{1\s}=\sqrt{2m_1^*(E_x+h_0\s)}/\hbar\  ,
\end{equation}
\begin{equation}
k_{3}=\sqrt{2m_3^*(E_x+eV_a+E_{\rm F,NM}-E_{\rm FL})}/\hbar\  ,
\end{equation}
\begin{equation}
k_{4\s}=\sqrt{2m_4^*(E_x+eV_a+\eta h_0\s)}/\hbar\  ,
\end{equation}
are the electron wave vectors along the $x$ axis. Here, $h_0=|{\bf h}_j|$ and
$\eta=+1 (-1)$ for parallel (antiparallel) alignment of the magnetizations.
The coefficients $A_{j\s}$ and $B_{j\s}$ are constants to be
determined from the boundary conditions, while
\begin{eqnarray}\label{rho}
\rho_{\s}(x)=\frac{x}{\lambda}+\beta_{\s}\ ,
\end{eqnarray}
with
\begin{equation}
\lambda=\left[-\frac{\hbar^2t_{\rm bar}}{2m^*_2eV_a}\right]^{1/3}\ ,
\end{equation}
\begin{equation}
\beta_{\s}=\frac{
[E_x-E_{\rm FL}-\phi_{\rm bar}-V^\s]t_{\rm bar}}{eV_a\lambda}\ .\\
\end{equation}
Although the transverse momentum
${\bf k}_\parallel$ is omitted from the above notations, the summation over
${\bf k}_\parallel$ is carried out in our calculations.

Upon applying the boundary conditions such that the wave functions
and their first derivatives are matched at each interface point
$x_j$ , i.e.  $\psi_{j,\s}(x_j)=\psi_{j+1,\s}(x_j)$ and
$(m^*_j)^{-1}[d\psi_{j,\s}(x_j)/dx]=(m^*_{j+1})^{-1}[d\psi_{j+1,\s}(x_j)/dx]$,
we obtain a matrix formula that connects the coefficients $A_{1\s}$ and
$B_{1\s}$ with the coefficients
$A_{4\s}$ and $B_{4\s}$ as follows:
\begin{equation}
\left[\begin{array}{c}
A_{1\s}\\B_{1\s}
\end{array}\right]
=M_{total}\left[\begin{array}{c}A_{4\s}\\B_{4\s}\end{array}\right],
\end{equation}
where
\begin{eqnarray}\label{M}
M_{total}&=&\frac{k_{4\s}}{k_{1\s}}
\left[\begin{array}{cc}
ik_{1\s}&\frac{1}{\lambda}\frac{m_1^*}{m_2^*}\\
ik_{1\s}&-\frac{1}{\lambda}\frac{m_1^*}{m_2^*}
\end{array}\right]
\left[\begin{array}{cc}
{\rm Ai}[\rho_{\s}(x=0)]&{\rm Bi}[\rho_{\s}(x=0)]\\
{\rm Ai}'[\rho_{\s}(x=0)]&{\rm Bi}'[\rho_{\s}(x=0)]
\end{array}\right]\nn\\&&
\times\left[\begin{array}{cc}
{\rm Ai}[\rho_{\s}(x=t_{\rm bar})]&{\rm Bi}[\rho_{\s}(x=t_{\rm bar})]\\
\frac{1}{\lambda}\frac{1}{m^*_2}{\rm Ai}'[\rho_{\s}(x=t_{\rm bar})]&
\frac{1}{\lambda}\frac{1}{m^*_2}{\rm Bi}'[\rho_{\s}(x=t_{\rm bar})]
\end{array}\right]^{-1}\nn \\&&
\times\left[\begin{array}{cc}
\cos(k_3t_{\rm NM})& -\frac{m^*_3}{k_3}\sin(k_3t_{\rm NM})\\
\frac{k_3}{m^*_3}\sin(k_3t_{\rm NM})& \cos(k_3t_{\rm NM})
\end{array}\right]
\left[\begin{array}{cc}
ik_{4\s}&m_4^*\\
ik_{4\s}&-m_4^*
\end{array}\right]^{-1}\nn\\&&
\times\left[\begin{array}{cc}
e^{-ik_{4\s}(t_{\rm bar}+t_{\rm NM})}&0\\
0&e^{ik_{4\s}(t_{\rm bar}+t_{\rm NM}})
\end{array}\right]^{-1}.
\end{eqnarray}
Since there is no reflection in region 4, the coefficient
$B_{4\s}$ in Eq. (\ref{psi}) is zero and the transmission
coefficient (TC) of the spin $\s$ electron, which is defined as the
ratio of the transmitted flux to the incident flux can be written as
\begin{equation}\label{T}
T_\s(E_x,V_a)=\frac{k_{4\s}m_1^*}{k_{1\s}m_4^*}\left|\frac{1}
{M^{11}_{total}}\right|^2,
\end{equation}
where $M^{11}_{total}$ is the left-upper element of the
matrix $M_{total}$ defined in Eq. (\ref{M}). At $T$=0 K, the
spin-dependent current density for the magnetic tunnel junction at
a given applied bias $V_a$ within the nearly-free-electron model is given
by the formula \cite{Duke}:
\begin{equation}\label{J}
J_\s(V_a)=\frac{em^*_1}{4\pi^2\hbar^3}\left[eV_a\int_{E_0^\s}^{E_{\rm
FL}-eV_a} T_\s(E_x,V_a)dE_x+\int_{E_{\rm FL}-eV_a}^{E_{\rm
FL}}(E_{\rm FL}-E_x) T_\s(E_x,V_a)dE_x\right],
\end{equation}
where $E^\s_0$ is the lowest possible energy that will allow transmission
and is given by $E^{\up}_0 =\max\{-h_0,-eV_a-\eta h_0\}$ for spin-up
and $E^{\dd}_{0} =h_0$ for spin-down electrons.
It is clear that the tunnel current is modulated by the magnetic configurations
of the both FM electrodes.

The degree of spin polarization for the tunnel current is defined by
\begin{equation}\label{p}
P=\frac{J_\up-J_\dd}{J_\up+J_\dd},
\end{equation}
where $J_\up$ $(J_\dd)$ is the spin-up (spin-down) current
density. For the present structure, we obtain this quantity when
the magnetizations of two FM electrodes and the FMS layer
(magnetic barrier) are in parallel alignment. For calculating the
TMR, both spin currents in the parallel and antiparallel
alignments are calculated. In this case, the TMR can be defined as
\begin{equation}\label{tmr}
\mbox{TMR}=\frac{(J_{\up}^{\rm P}+J_{\dd}^{\rm P})-(J_{\up}^{\rm AP}
+J_{\dd}^{\rm AP})}{J_{\up}^{\rm AP}+J_{\dd}^{\rm AP}},
\end{equation}
where $J_{\up,\dd}^{\rm P(AP)}$ corresponds to the current density in the
parallel (antiparallel) alignments of the magnetizations in the FM
electrodes, for a spin-up ($\up$) or spin-down ($\dd$) electron.

\section{Numerical results and discussion}
Numerical calculations have been carried out to investigate the
effect of a magnetic barrier and an NM metallic layer on the spin
transport in the Fe/EuS/Au/Fe structure as a typical MTJ. We have
chosen Fe and EuS because they have cubic structures and the
lattice mismatch is very small \cite{Dem}. The parameters $E_{\rm
F}$ and $h_0$ for Fe electrodes are taken corresponding to $k_{\rm
F\up}$=1.09 \AA$^{-1}$ and $k_{\rm F\dd}$=0.42 \AA$^{-1}$ (for
itinerant $d$ electrons) \cite{Stear}. For the EuS layer we use
$S$=7/2, $I$=0.1 eV \cite{Nolt}, $\phi$=1.94 eV \cite{Saffar2} as
a symmetric barrier height, and $t_{\rm bar}$=1.3788 nm which
corresponds to four monolayers of EuS$\las 111\ras$ \cite{Maug}.
The Fermi energy in the Au layer is $E_{\rm F,NM}=5.51$ eV
\cite{Kittel}. In practice, the effective masses of electrons may
differ from that of free electron, but here for simplicity, we
assume all electrons have the same mass, $m_e$ as free electrons.

In our considered system, the magnetization direction of the left Fe
electrode and the EuS magnetic barrier stays fixed, because they are
coupled through the interface. Inserting the NM spacer layer between
the FMS layer and the right FM electrode decreases the exchange coupling
between the two FM electrodes. Therefore the right Fe electrode is free
and may be switched back and forth by an external magnetic field
(see Fig. 1). In the following, we show the numerical results at $T$=0 K,
so in the magnetic barrier we have $\las S_z\ras=S$.

The TC for tunneling electrons at the Fermi energy, and the TMR in
the tunnel junction as a function of the NM layer thickness are
shown in Figs. 2(a) and 2, respectively. It can be seen that both
the TC and the TMR oscillate with the NM layer thickness due to
the dependence of these quantities on the incident electron's
energy and the NM layer thickness. The physics behind this
behavior is easy to understand. When the energy of the incident
electrons is approximately equal to the energy of a quasibound
state in the quantum well, a resonance condition is fulfilled and
the TC of the electrons through the structure strongly increases.
On the other hand, the position of the quantum well levels, formed
in the NM layer, strongly depends on the well thickness.
Therefore, with continuous variation of the $t_{\rm NM}$, the
position of the resonant states varies and this leads to the
oscillations of the TC and the TMR. From Fig. 2, it is clear that
where the transmission is high, the TMR is low.  From the
continuous variation of $t_{\rm NM}$, one can see that both the TC
and the TMR oscillate with a period of 0.26 nm which is in
excellent agreement with $\lambda_F$/2, where $\lambda_F$ is the
Fermi wavelength in the Au layer and is equal to 0.522 nm. The
results show that, for the spin-up electrons, the TC is very
higher than the spin-down ones both in the parallel and
antiparallel alignment of magnetizations. The reason can be
explained by the spin filtering effect of the FMS barrier. As
shown in Fig. 1, due to the spin splitting of the conduction band
in the FMS layer, the barrier height for spin-up electrons is
lowered and the tunneling probability greatly increases, while for
the spin-down electrons the barrier height is raised and hence,
the tunneling probability for these electrons reduces. Thus, a
large spin-polarized current is produced.

In Fig. 2(b) we have also shown the TMR in terms of thickness, when
$t_{\rm NM}$ changes in monolayer steps. At the monolayer thickness
$t_{\rm ML,NM}$=0.2355 nm, which corresponds to interlayer distance of
Au$\las111\ras$, the TMR first increases in the first three monolayer and
after that the TMR becomes less and starts to oscillate with a period of
approximately 2.5 nm. In contrast, for $t_{\rm ML,NM}$=0.27 nm, the TMR
initially decreases and the period of oscillations is approximately 7 nm.

The voltage dependence of the TMR for several thicknesses is shown in Fig. 3,
when $t_{\rm ML,NM}$=0.2355 nm. We can see that, due to the FMS layer, the TMR
is asymmetric, as a tunnel junction without a nonmagnetic layer is used. For
each fixed voltage, the TMR oscillates, as is shown in Fig. 2(b) for
$V_a$=50 mV. The results show that more than 250\% TMR can be obtained in
reverse bias. By increasing the number of nonmagnetic monolayers, the TMR
shows an oscillatory behavior in the positive voltages. Under a forward bias,
electrons at the Fermi energy of the left electrode, tunnel into the empty
states of the quantum well, which have been created due to the applied bias.
Therefore the applied bias has a strong influence on the TMR and the
oscillatory behavior will appear. On the other hand, under a reverse bias,
electrons at the Fermi energy of the quantum well tunnel into the empty states
of the left electrode. In this case, variations of the applied bias cannot
strongly affect the TMR.

In order to further see the effect of the FMS layer on the tunnel
currents, we have shown in Fig. 4 the spin polarization of
tunneling electrons as a function of the bias voltage, for $t_{\rm
ML,NM}$=0.2355 nm. It is obvious that the spin polarization is
symmetric for $t_{\rm ML,NM}$=0, and becomes strongly asymmetric
when the number of monolayers increases. The results show that the
spin polarization can reach 99\%, which is evidence of spin
filtering effect in the FMS layer.

The experimental measurements show that, if one uses a nonmagnetic
insulator instead of a magnetic tunnel barrier, very low values
for electron-spin polarization and TMR will result
\cite{Mood2,Yuasa}. The oscillatory behavior of TMR in terms of
the NM layer thickness, is qualitatively in agreement with the
experiments of Yuasa {\it et al} \cite{Yuasa}, and shows that the
coherent tunneling of electrons through the quantum well states is
the main reason of this behavior.

The present results can be qualitatively compared with those
already known in the relevant literature, when a nonmagnetic
barrier is used. In \cite{Ved,WilBar}, the authors showed that in
the FM/NM/I/NM/FM systems, the TMR in two different cases has an
oscillatory behavior with NM layers: the case where the thickness
of both NM layers are equally varied (symmetric case), and the
case where only one of the NM layers is increased (asymmetric
case). In the symmetric case, the TMR is always positive, but in
the asymmetric case the TMR oscillates from positive to negative
as a function of the NM layer thickness. The situation of the
asymmetric case is similar to our structure in which only one NM
layer has been used \cite{Zhang2}. Thus, as it is clear from Figs.
2(b) and 3, our results are in qualitative agreement with the
asymmetric case in FM/NM/I/NM/FM structures.

Finally, we discuss the effect of interface scattering on
spin-injection from the FM into NM spacer \cite{Blon,Hu}. For this
purpose, we treated the magnetic barrier as a $\delta$-function
$U_\s\delta(x)$, with $U_\s$ describing the barrier strength
parameter for spin $\s$ electron. In this case, the parameter
$U_\s$ is proportional to both height and width of the barrier.
The numerical results showed that with increasing the $U_\s$, the
TC reduces, and hence the spin-dependent currents decay
exponentially, which is due to the interface resistance. On the
other hand, the spin polarization and TMR increase with $U_\s$,
and trend gradually towards constant values, because in these
values of the barrier parameter, only the FM electrodes have a
dominant influence on the spin polarization and TMR.

\section{Concluding remarks}
The magnetotransport behavior in an FM/FMS/NM/FM tunnel junctions
was investigated theoretically, using the transfer matrix method
and the nearly-free-electron approximation. The numerical results
indicate that in Fe/EuS/Au/Fe structure, the TMR has an
oscillatory behavior as a function of the NM layer thickness. By
increasing the thickness of the NM layer, both the TMR and
electron-spin polarization, oscillate under the forward biases.
Due to the presence of a magnetic barrier, there exist more than
250\% TMR and 98\% spin polarization in the tunnel currents; this
can be obtained by adjusting the thickness and the applied bias.

Our study restricted to the zero temperature limit, because the ferromagnetic
transition temperature in most of the FMS layers is much lower than room
temperature \cite{Maug}. Low temperature spin transport is, however, important
as a testing ground for novel ideas and concepts in future spin electronic
devices based on coherent transport \cite{Yuasa,Dietl}.

\section*{Acknowledgments} One of us (A.A.S.) would like to
express his sincere thanks to Prof. K. Esfarjani for illuminative
comments.

\begin{figure}
\begin{center}
\leavevmode\hbox{\epsfxsize=1.15\textwidth\epsffile{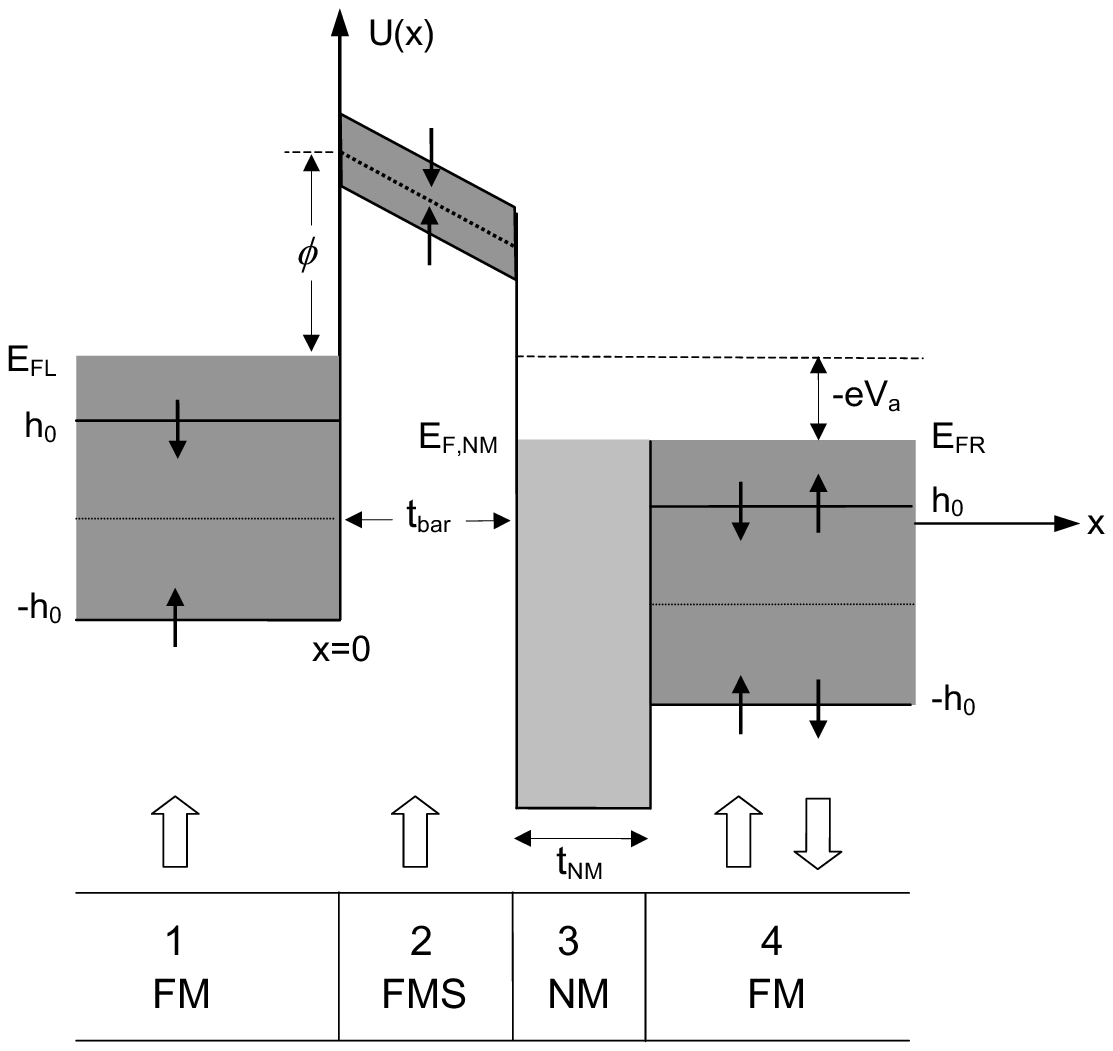}}
\caption{Spin-dependent potential profile for an FM/FMS/NM/FM
magnetic tunnel junction under forward bias $V_a$. In the FMS
layer, the dashed line represents the bottom of the conduction
band at $T\geq T_C$ and the thin arrows indicate the bottom of the
conduction band for spin-up and spin-down electrons at $T<T_C$.
The zero of energy is taken at the middle of the bottoms for the
majority-spin band and minority-spin one in the left FM
electrode.}
\end{center}
\end{figure}

\begin{figure}
\begin{center}
\leavevmode\hbox{\epsfxsize=0.8\textwidth\epsffile{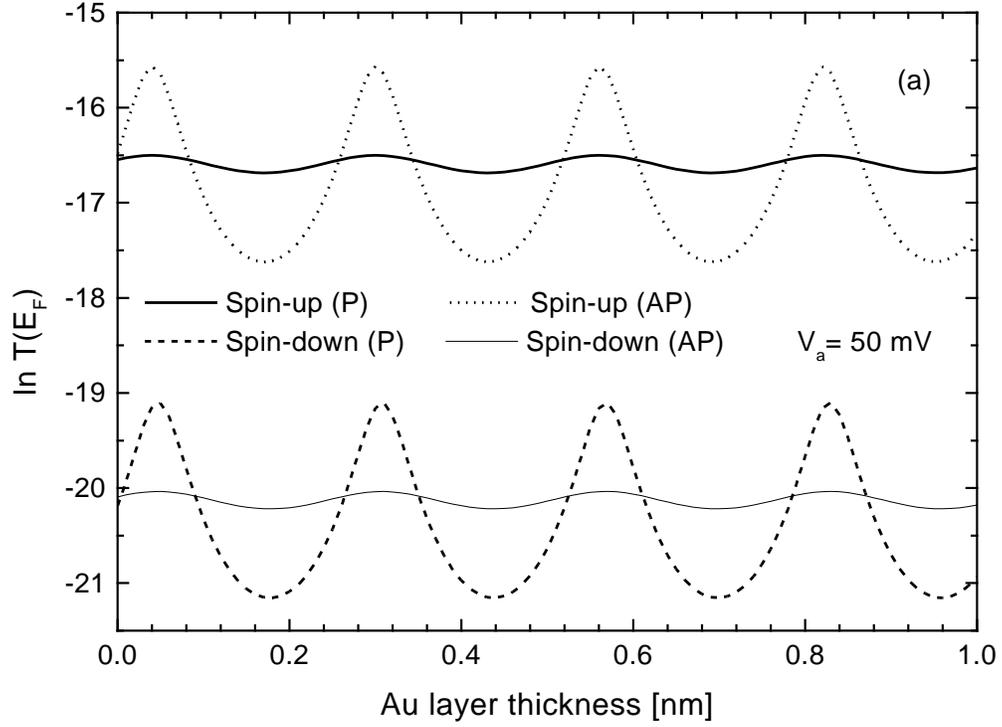}}
\leavevmode\hbox{\epsfxsize=0.8\textwidth\epsffile{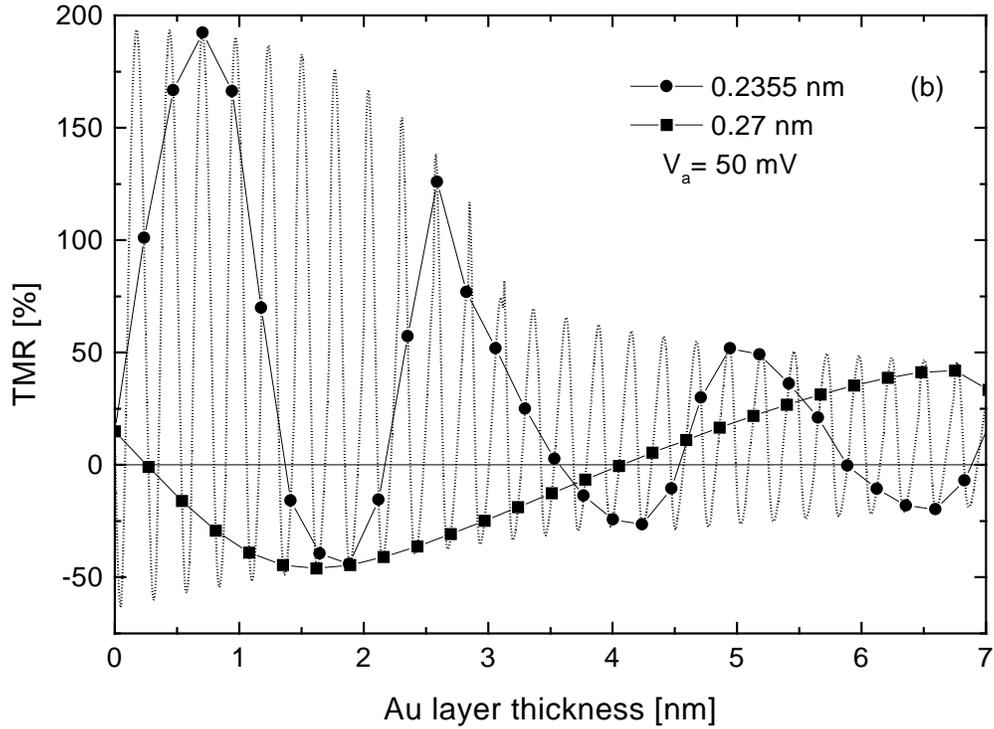}}
\caption{(a) Spin-dependent TC at Fermi energy in the parallel (P)
and antiparallel (AP) configurations, and (b) dependence of the
TMR, as a function of Au layer thickness.}
\end{center}
\end{figure}

\begin{figure}
\begin{center}
\leavevmode\hbox{\epsfxsize=0.8\textwidth\epsffile{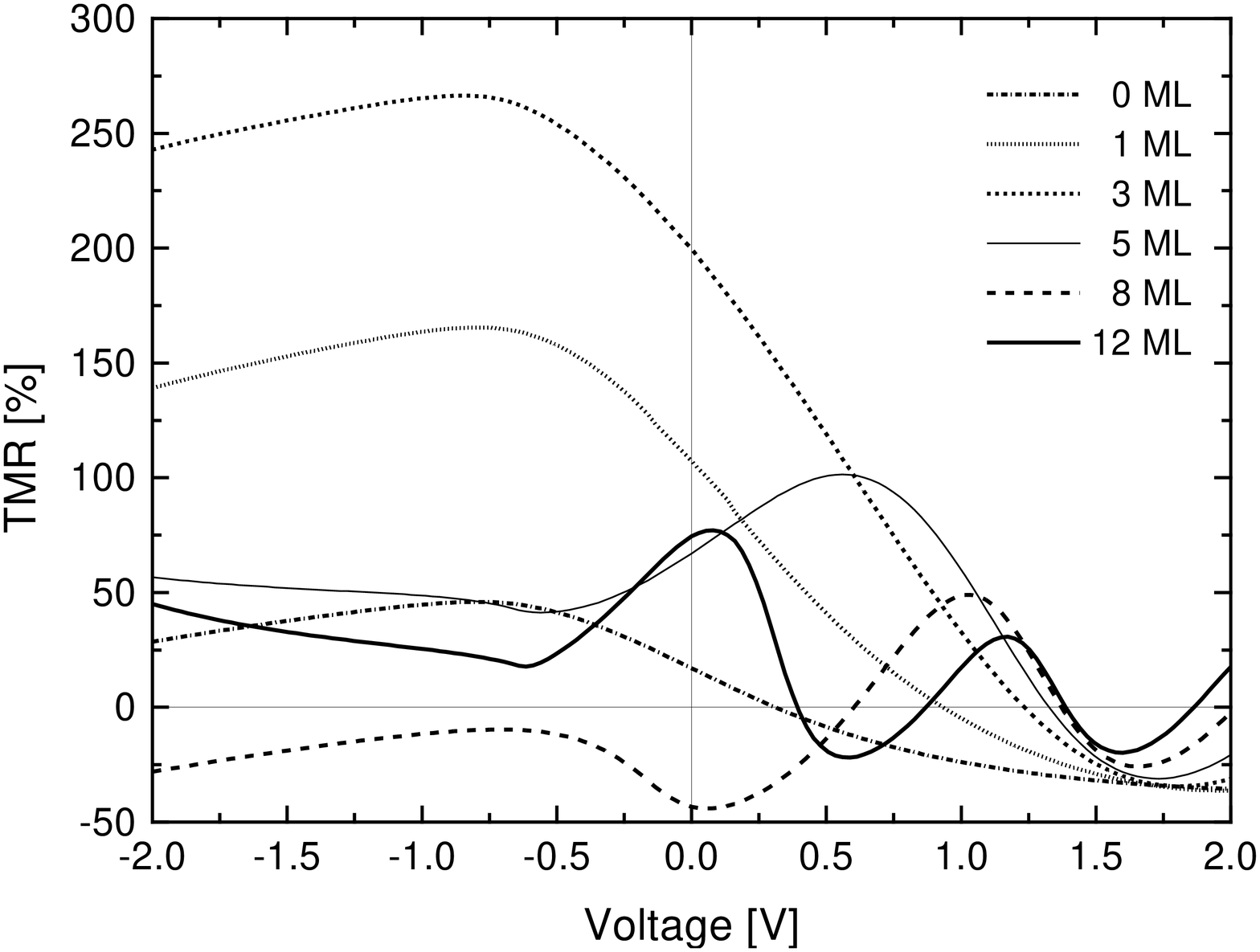}}
\caption{Dependence of the TMR on the applied voltage for
different NM multilayers.}
\end{center}
\end{figure}

\begin{figure}
\begin{center}
\leavevmode\hbox{\epsfxsize=0.8\textwidth\epsffile{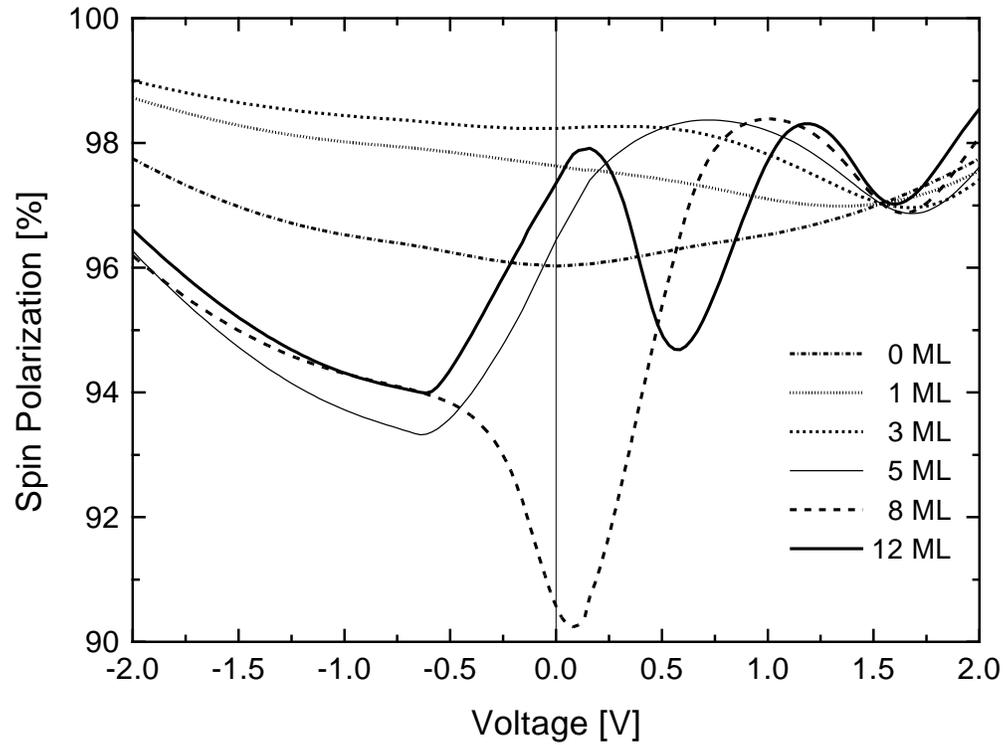}}
\caption{Dependence of the spin polarization on the applied
voltage for different NM multilayers.}
\end{center}
\end{figure}

\end{document}